\documentclass{ws-ijmpa}
\usepackage{subfigure}
\begin{document}

\markboth{L. Le\'sniak}
{New parameterization of the resonant production amplitudes near an inelastic
threshold}

%
\catchline{}{}{}{}{}
%
\newcommand{\kk}{$\rm K \overline{K}$ }
\newcommand{\be}{\begin{equation}}
\newcommand{\ee}{\end{equation}}

\title{\bf NEW PARAMETERIZATION OF THE RESONANT PRODUCTION AMPLITUDES NEAR AN INELASTIC
  THRESHOLD}

\author{\footnotesize L. LE\'SNIAK$^1$}
\vspace{0.5cm}
\address{\it \small  $^1$ Division of Theoretical Physics, The Henryk Niewodnicza\'nski 
Institute of Nuclear Physics, Polish Academy of Sciences, 31-342 Krak\'ow,
 Poland \\ Leonard.Lesniak@ifj.edu.pl\\}
 
\maketitle


\begin{abstract}New formulae for the resonant scattering and the production 
amplitudes near an inelastic threshold are derived. It is shown that the 
Flatt\'e formula, frequently used in experimental analyses, is not sufficiently
accurate. Its application to data analysis can lead to a substantial distortion of the effective
mass spectra and of the resonance pole positions.

A unitary parameterization, satisfying a generalized Watson theorem for
the production amplitudes, is proposed. It can be easily applied to study
production processes, multichannel meson-meson interactions and the resonance 
properties, including among others the scalar resonances $ a_0(980)$ and 
$f_0(980)$.


\end{abstract}

\ccode{PACS numbers: 11.80.Gw, 11.80.Fw, 13.75Lb}

\section{\bf Introduction }

Resonant scattering amplitudes near inelastic thresholds cannot be accurately 
described
by the Breit-Wigner formula. About 30 years ago S. M. Flatt\'e proposed the 
following parameterization of the $S$-wave $\pi\eta$ and the \kk production 
amplitudes which are dominated by the $a_0(980)$ resonance \cite{Flatte}:
 
 \be 
A_j \sim \frac{M_R \sqrt{\Gamma_0
\Gamma_j}}{M_R^2-E^2-iM_R(\Gamma_1+\Gamma_2)},~~~j=1,2. \label{Flatte}
\ee 
Here $M_R$ is the resonance mass, $E$ is the effective mass (c.m. energy) and
$\Gamma_j=g_j k_j$ are the channel widths, $k_1$ is the pion (or $\eta$)
c.m. momentum, $k_2$ is the kaon c.m. momentum and $g_i$ are the channel
coupling constants. Below the \kk threshold $\Gamma_2$ is imaginary:
$\Gamma_2= i g_2 p_2$, where $ p_2=\sqrt{m_K^2-\frac{E^2}{4}}$, $m_K$ being the
kaon mass. At the \kk threshold the energy $E_0=2 m_K$, $k_1=q$ and
$\Gamma_0=g_1 q$. Above the threshold $E=2 (k_2^2+m_K^2)^{\frac{1}{2}}$. 

From (\ref{Flatte}) we see that apart of the normalization factors the
Flatt\'{e} production amplitudes (1) depend on three real
parameters: the resonance mass $M_R$ and two coupling 
constants $g_1$ and $g_2$. One can, however, easily demonstrate that in presence
of an inelastic coupling to the $\pi \eta$ channel the three
parameters are not sufficient to describe a behaviour of the \kk elastic 
scattering amplitude $T_{22}$ near its threshold. If this coupling is switched 
off then the
following threshold formula, called the effective range approximation, can be
used: 
\be
T_{22}=\frac{1}{k_2 \cot \delta_2 -i k_2},~~~~~~~~~~~~
k_2 \cot \delta_2 \approx\frac{1}{a} +\frac{1}{2} r~ k_2^2.   \label{eff1}
\ee 
Here $\delta_2$ is the \kk phase shift and the two real parameters $a$ and $r$ 
denote the \kk scattering length and the effective range, respectively.
However, if the interchannel coupling is nonzero then one has to introduce the
inelasticity parameter $\eta$ and to replace the real parameters $a$ and $r$ by 
the complex ones, $A$ and $R$:
\be
T_{22}=\frac{1}{2ik_2} ( \eta e^{2 i \delta_2} - 1) \approx
\frac{1}{\frac{1}{A} - i~ k_2 +\frac{1}{2}~ R~ k_2^2}.  \label{eff2}
\ee 
It means that one needs at least four real parameters to describe the \kk
scattering near its threshold. These parameters should also appear in the
formulae for the production amplitudes $A_1$ and $A_2$. This can be achieved
by introduction of a new complex constant $N$ in the denominator $W$ of the
production amplitudes $A_i$ in (\ref{Flatte}): 
\be
W= M_R^2 - E^2-i M_R g_1 q -i M_R g_2 k_2 +N~k_2^2.  \label{Wu}                                                     
\ee
Dividing $W$ by the product $M_R g_2$ we get
\be
\frac{W}{M_R g_2}= \frac{1}{A} - i k_2 +\frac{1}{2} R k_2^2,   \label{WMR}
\ee
where the inverse of the scattering length $A$ and the effective range $R$ are
written in terms of the three Flatt\'e parameters $M_R, g_1$, $g_2$ and of the
new parameter $N$:
\be
\frac{1}{A}=\frac{M_R^2 - E_0^2}{M_R g_2}-i\frac{g_1}{g_2}~q,~~~~~~~~~
R=\frac{2 N-8}{M_R g_2}.  \label{AR}
\ee
In the Flatt\'{e} approximation $N=0$, hence  $Re R=-8/M_R g_2$ 
and $Im R \equiv 0$. The zero value of the imaginary part of the effective range is an
essential limitation of the Flatt\'{e} formula. 

\section{\bf Scattering Amplitudes}

The first channel elastic scattering amplitude $T_{11}$ can be parameterized in 
terms of
five parameters. The four of them: $Re A, Im A, Re R$ and $Im R$ are the same 
as those defined in (\ref{eff2}) for the second channel amplitude. The fifth
parameter is the first channel phase shift $\delta_0$ determined at the \kk
threshold. Using the two-channel unitarity one can derive the following
approximate formula for $T_{11}$:
 \be
T_{11} \approx \frac{e^{i\delta_0}}{k_1} \frac{\sin \delta_0~ +~i~
Im~(e^{-i\delta_0}~A)~k_2~-~\frac{1}{2}~Im~(e^{-i\delta_0}~A~R)~
k_2^2}{1-~i~ A~k_2~+~\frac{1}{2}~A~R ~k_2^2}\cdot     \label{T11new}
\ee
Below the \kk threshold one has to replace $k_2$ by $ip_2$. In the Flatt\'{e} 
approximation $\delta_0$ equals to the  phase of the complex
scattering length $A$. 

The transition amplitude $T_{12}$ from the first to the second channel in the
new approach is given by
\be
T_{12} \approx \frac{1}{\sqrt{k_1}}~ e^{i \delta_0} \frac{\sqrt{Im~ A~ -~\frac{1}{2}~ |A|^2
~Im~R~ k_2^2}}{1-~i~ A~k_2~+~\frac{1}{2}~A~R ~k_2^2}\cdot  \label{T12new} 
\ee

Let us remark that all the three amplitudes, given by Eqs. (\ref{eff2}), 
(\ref{T11new}) and (\ref{T12new}), have a common denominator
\be
D(k_2)=1-~i~ A~k_2~+~
\frac{1}{2}~A~R ~k_2^2 \cdot             \label{D}
\ee
The complex zeroes $z_{1,2}$ of $D(k_2)$ are related to $A$ and $R$ by    
\be
z_{1,2} = \frac{i}{R} \pm \sqrt{-\frac{1}{R^2}-\frac{2}{AR}}\cdot \label{z1z2}
\ee
In the Flatt\'{e} approximation $Re~z_1~=- Re~z_2$ which limits possible positions
of the amplitude poles at the energies  $E_{1,2}=\sqrt{E_0^2+4z_{1,2}^2}$. 
These pole positions are the most essential quantities characterizing the
resonances.
\begin{figure}[pb]
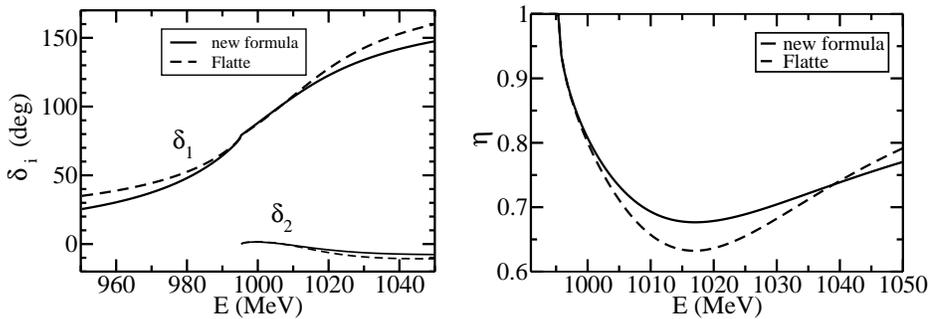

 \includegraphics[height=.213\textheight]{delta12cz.eps}~~~ 
  \includegraphics[angle=0,height=.215\textheight]{eta.eps}   
\vspace*{8pt}
\caption{Comparison of phase shifts and inelasticities of the scattering amplitudes\label{f1}}
\end{figure}


\section{\bf Phenomenological Studies of the $a_0(980)$ Threshold Parameters}

It is instructive to show numerical results concerning the threshold parameters
of the $a_0(980)$ resonance, situated close to the \kk threshold. One can use a 
coupled channel formalism for the separable meson-meson interactions in two or
three channels. It  has been developed in Ref.~\refcite{LL} and further
 applied to study the $a_0$ resonances
in the $\pi \eta$ and the \kk channels \cite{AFLL}. The model parameters were 
fixed using 
the data of the Crystal Barrel and of the E852 Collaborations. The 
following threshold parameters have been presently calculated:
$Re~ A= 0.17$ fm, $Im ~ A=0.41$ fm, $Re ~R = -11.32$ fm, and $Im ~R = -3.18$  fm.
Let us stress here that the imaginary part of the effective range is nonzero and
cannot be neglected. In Fig. 1 we see differences between the phase
shifts and inelasticities calculated using the theoretical model of 
Ref.~\refcite{AFLL} and the Flatt\'e approximation in which $Im R \equiv 0$. They are 
sufficiently large to
create deviations of the order of hundred percent between the squares of the 
moduli of the scattering amplitudes $|T_{11}|^2$ or $|T_{22}|^2$ already at the 
distance of 50 MeV away from the \kk threshold (see Fig. 1 of Ref.~\refcite{Lizbona}).
Another discrepancy between the Flatt\'e formula and the above model is a shift
of the pole complex energies $E_1$ and $E_2$. For example, the $Re E_1$ shift is 
larger than 10 MeV and exceeds the experimental energy resolution of
many present experiments (see Fig. 2 of Ref.~\refcite{Lizbona}).

\section{\bf New Formulae for the Production Amplitudes}

Let us assume that there are no initial state strong interactions. Then one can
generalize the Watson theorem, which is satisfied below the inelastic threshold
\be
Im~ A_1 = k_1~ T_{11}~ A_1^*~,  \label{Watson}
\ee
to a form valid for the two coupled channels above this threshold:
\begin{eqnarray}
 Im~ A_1= k_1~ T_{11}~ A_1^*~+k_2~ T_{12}~ A_2^*,\\
 Im~ A_2= k_2~ T_{22}~ A_2^*~+~k_1 ~T_{21}~ A_1^*. \label{Watson2}
\end{eqnarray}
One can propose the following new parameterization of the production
amplitudes:
\be
A_1= f_1~T_{11}~+~f_2~T_{12},~~~~~A_2= f_1~T_{12}~+~f_2~T_{22}~. \label{A1A2} 
\ee
Here $f_1$, $f_2$ are real functions of energy (or momentum $k_2$) which near
the threshold can be approximated by 
\be
 f_1 \approx n_1+\beta_1 k_2^2,~~~
 f_2 \approx n_2+\beta_2 k_2^2.   \label{n1n2}
\ee
In (\ref{n1n2}) $n_1$ and $n_2$ are normalization constants, $\beta_1$ and 
$\beta_2$
are additional real coefficients. The formula given by (\ref{Flatte}) is finally
replaced by (\ref{A1A2}). The parameters to be fitted from experiments are:
complex $A,R$ and real $\delta_0,n_1,n_2,\beta_1,\beta_2$.

 A generalization of
the above formulae to a case where the particle masses in the second
channel are different ($m_a$ and $m_b$) is simple. Above the inelastic 
threshold one defines the
momentum   $ k_2=\frac{1}{2E}
\{[E^2-(m_{a}+m_{b})^2][E^2-(m_{a}-m_{b})^2]\}^{\frac{1}{2}}$.
Below the threshold (for $E< E_0=m_{a}+m_{b}$) $k_2$ is
replaced by $i p_2$, where $p_2=\frac{1}{2E} \{[(m_{a}+m_{b})^2-E^2]
[E^2-(m_{a}-m_{b})^2]\}^{\frac{1}{2}}$. 
 
The new formulae can be applied in numerous analyses of present and future
experiments (for example Belle, BaBar, CLEO, BES, KLOE, COSY, Tevatron, LHCb,
JLab, PANDA ...) and also to reanalyse older experiments in order to update
our information on meson spectroscopy and on reaction mechanisms.\\



\begin{thebibliography}{0}    

\bibitem{Flatte}
S. M. Flatt\'e, {\it  Phys. Lett.} \textbf{B63}, 224 (1976).

\bibitem{LL}
L. Le\'sniak, {\it  Acta Physica Polonica} \textbf{B27}, 1835 (1996).
 
\bibitem{AFLL}
Agnieszka Furman and Leonard Le\'sniak, {\it Phys. Lett.} \textbf{B538}, 266 (2002).
 
\bibitem{Lizbona} L. Le\'sniak, AIP Conf. Proc. \textbf{1030}, 238 (2008),
 arXiv:0804.3479 [hep-ph].

\end{thebibliography}
\end{document}